\documentclass[aps,pra,superscriptaddress,tightenlines,twocolumn,showpacs,10pt,floatfix]{revtex4-1}

\usepackage{amssymb,amsmath}
\usepackage{graphicx}
\usepackage{eurosym}
\usepackage{color}
\usepackage[normalem]{ulem}
\usepackage[makeroom]{cancel}
\usepackage[none]{hyphenat}
\usepackage{hyperref}
\usepackage[hyphenbreaks]{breakurl}
\hypersetup{backref,colorlinks=true,linkcolor=blue,citecolor=blue,urlcolor=blue,linktocpage=true,breaklinks}

\newcommand{\bn}{{\bf n}}
\newcommand{\bE}{{\bf E}}
\newcommand{\bH}{{\bf H}}

\newcommand{\teta}{\tilde{\eta}}

\newcommand{\tQ}{\tilde{Q}}
\newcommand{\tq}{\tilde{q}}
\newcommand{\TE}{{\rm TE}}
\newcommand{\TM}{{\rm TM}}

\begin{document}
\title{Modes and exceptional points in waveguides with impedance boundary conditions}
\author{Bikashkali Midya} 
\email{bmidya@ist.ac.at}
\affiliation{Institute of Science and Technology Austria, Am Campus 1, 3400 Klosterneuburg, Austria}
\author{Vladimir V. Konotop}
\email{vvkonotop@fc.ul.pt }
 \affiliation{Centro de Fisica Te\'orica e Computacional and Departamento de F\'isica, Faculdade de Ci\^encias,
Universidade de Lisboa,  Campo Grande 2, Edif\'icio C8, Lisboa 1749-016, Portugal}
 




\begin{abstract} 
A planar waveguide with impedance boundary, composed of non-perfect metallic plates,  and with passive or active dielectric filling is considered. We show the possibility of selective mode guiding and amplification when homogeneous pump is added to the dielectric, and analyze differences in TE and TM mode propagation. Such a non-conservative system is also shown to feature exceptional points, for specific and experimentally tunable parameters, which are described for a particular case of transparent dielectric.
\end{abstract}


\maketitle

A localized gain can support stationary mode propagation in guiding structures. This is a long-standing subject~\cite{semiconduct}, which recently received considerable attention in optics~\cite{Siegman} and plasmonics~\cite{plasmonic}, where the gain guidance was observed experimentally~\cite{SiegAtAll,SudeshEtAll}.  Since media with gain and dissipation are characterized by complex refractive indexes, the eigenmode structures of the respective waveguides are determined by solving non-Hermitian eigenvalue problems. Such problems generally have complex eigenvalues, and thus modes {are} characterized by either amplification or by absorption. The guidance, therefore, becomes feasible only due to the eventual possibility of a non-Hermitian operator having real eigenvalues.  Remarkably, if in a guiding medium the gain co-exists in a delicate balance with absorption, it may happen that all eigenvalues are real {(below a certain critical value of the gain parameter)}. In some situations the latter is possible only if the medium obeys special symmetries~(see e.g.~\cite{review} for a recent review) like, for example the parity-time symmetry~\cite{BenderBoet}. In those cases all modes are guided without amplification or absorption~\cite{OL_Christodol}. This is a property of not only linear systems: gain guidance is also possible in the nonlinear regime when localized gain is present in a generically absorbing Kerr medium~\cite{ZKK} (see also~\cite{KarKonZez}) where a stable mode (if any) propagates in a form of a dissipative soliton~\cite{Akhmeddiev}. 

Wave guidance in non-conservative media is of practical interest, what is justified by a number of features of such systems not available in the conservative setting. Indeed, gain can be implemented in media with ubiquitous natural losses (especially strong in plasmonics where metallic elements are present) suing say active impurities~\cite{Kipp}. Such gain is induced by an external pump field and thus allows for changing guidance properties {\em in situ}. Moreover, non-Hermitian systems allow for the existence of exceptional points~\cite{except,review} (EPs) at which the behavior of the system (or of several modes) is changed qualitatively. This opens novel possibilities on the control over the system, having no analogs in the conservative propagation. In particular, we mention mode exchange~\cite{DGHHHRR}, suppression of the decay~\cite{Zeno}, and even revival of lasing~\cite{NatPhys} induced by increasing dissipation.    

If a linear waveguide is characterized by homogeneous absorption or pump, the latter has trivial effect on the modes: all of them undergoing decay or amplification and in most settings. This represents a net effect which can be scaled out by the proper exponential factor. In such situations, absorption and gain do not offer new possibilities for manipulating mode structure of a waveguide.  
\begin{figure}[ht!]
\centering
\includegraphics[width=0.9\columnwidth]{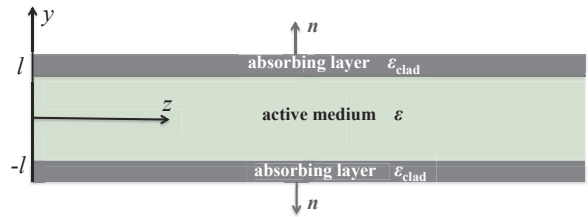}
\caption{Schematic presentation of the waveguide with impedance boundary. The field is polarized along $x$-direction.}
\label{fig:one}
\end{figure}

In the present Letter, we explore an alternative mechanism of guiding, amplifying, and managing modes in (active and passive) waveguides by considering absorption at the waveguide boundaries. {Such a waveguide might be favorable in practical setup because, on the one hand boundaries provide the most direct way of accessing and controlling the waveguide properties, and on the other hand the mode control is carried by the simplest homogeneous distributions of losses and gain.}   Like in majority of  dissipative structures, different modes in waveguides with absorbing boundaries are non-orthogonal and display different decrements. Thus, by applying gain inside the guiding domain one can amplify modes and even obtain guided modes selectively. Such waveguides also allow  observation of the EPs (similar finding was recently reported for acoustical waveguides with impedance boundary conditions~\cite{acoustics}) for specific amplitudes of the gain or absorption which remain homogeneous along the waveguide.

 Here, we consider waveguides of the type illustrated schematically in Fig.~\ref{fig:one}. An active (or passive) medium is bounded by two imperfect planar conductors located in the planes $y=\pm\ell$, creating a two-dimensional planar waveguide of the width $2\ell$. We consider monochromatic plane waves propagating along $z-$direction and independent on $x-$coordinate. For the sake of simplicity we assume that the dielectric between the conducting plates has a constant permittivity, $\epsilon$, which can be either real or complex.   
 
We constrain the choice of the boundary layers by their absorbing properties which can be approximately described by the impedance boundary conditions~\cite{Senior,LL}: 
\begin{equation}
{\bf n}\times{\bf E}=\eta{\bf H}. 
\label{BC}
\end{equation}
 Here $\bn$ is the normal to a given boundary directed outwards the guiding medium [see Fig.~\ref{fig:one}] and $\eta=\sqrt{\mu_{\rm clad}/\epsilon_{\rm clad}}$ is the surface impedance with $\mu_{\rm clad}$ being the permeability and $\epsilon_{\rm clad}$ being the complex  permittivity of the cladding. In the limit of $\eta =0$, our model corresponds to the conventional waveguide with perfect conducting boundary. We consider that the surface impedance has a positive real part, Re $\eta>0$, what corresponds to the boundary absorption~\cite{LL} and ensures the causality principle~\cite{Agarwal:book}. Restricting present consideration to nonmagnetic cladding, i.e. setting $\mu_{clad}=1$, below we discuss only the case where Im $\eta<0$ (this corresponds to positive conductivity of the metallic cladding).
  
 Important to mention that the impedance boundary conditions, first introduced in~\cite{Leontovich} for modeling propagation of radio-waves, are valid for surfaces characterized by sufficiently large $|\epsilon_{\rm clad}|^{1/2} \gg 1$ (see~\cite{Senior} for the derivation) and generically are not applicable for typical metals in the optical diapason~\cite{LL}. In the infrared radiation, however, several metals satisfy the required conditions. In particular, a sufficiently small parameter $|\epsilon_{\rm clad}|^{-1/2}$ defining the accuracy of the approach at wavelengths of order of $1.5$ $\mu$m with the required signs of the real and imaginary parts of the permittivity can be achieved in (see e.g. Ref.~\cite{Ordal}): Al ($\approx 0.06$ at 1.55 $\mu$m),  Au ($\approx 0.096$ at 1.5 $\mu$m), Pt ($\approx 0.11$ at 1.55 $\mu$m), and Ni ($\approx 0.13$ at 1.55 $\mu$m).  Impedance boundary conditions in optical systems were previously addressed in  studies of scattering by randomly varying smooth~\cite{Depine}, rough~\cite{Maradudin} and inhomogeneous~\cite{Dykhne} metallic surfaces. The cladding can be further brought closer to the "impedance regime" by increasing its absorbing characteristics by means of resonant impurities or by using metasurfaces~\cite{metasurface}. 

Below we consider equal cladding at $y=\pm\ell$  (generalization to different impedances is straightforward). Then the impedance boundary conditions 
are rewritten now as  
{$ E_x (\pm \ell)=\mp\eta H_z(\pm \ell)$ and
$ E_z(\pm \ell)=\pm\eta H_x(\pm \ell)$}. Inside the waveguide both electric, $\bE$, and magnetic, $\bH$ fields solve the Helmholtz equation, i.e. 
$ 
\nabla^2 \bE + k_0^2\epsilon  \bE =0,$ and 
$ \nabla^2 \bH +k_0^2\epsilon \bH =0 $, where 
$k_0=\omega/c = 2\pi/\lambda$, $\omega$ is the frequency and $\lambda$ is the wavelength in the vacuum, and {$\epsilon = \epsilon_r + i \epsilon_i$} is the permittivity of the guiding medium. 

TE and TM modes inside the  waveguide are searched in the form $\bE=(e^{i q z}\phi^{\rm TE}(y),0,0)$ and $\bH=(e^{i q z}\phi^{\rm TM}(y),0,0)$, where  $\phi^{\alpha}(y)$ with  $\alpha$ standing for either TE or TM,  describes the transverse field distributions.  Using dimensionless variable $Y=y/\ell$, one obtains two non-Hermitian Sturm-Liouville problems:  
\begin{subequations}
	\label{SL} 	 
	\begin{equation}
	\label{SL_TE}
	\phi_{YY}^{\TE} =- Q^2 \phi^{\TE},  \quad	\phi^{\rm TE}(\pm 1) = \pm \eta^{\rm TE}\phi_Y^\TE(\pm 1) 
	\end{equation}
for TE modes and
	\begin{equation}
	\label{SL_TM}
	\phi_{YY}^{\TM} =- Q^2 \phi^{\TM},  \quad	\phi_Y^{\rm TM}(\pm 1) = \pm \eta^{\rm TM}\phi^\TM(\pm 1), 
	\end{equation}
\end{subequations}
for TM modes. Here  $\phi_Y^\alpha \equiv d\phi^\alpha/dY$,  the spectral parameter $Q$ is related to the propagation constant $q$ by the equation
 \begin{eqnarray}
 \label{Q}
q^2 = \ell^2 k_0^2 \epsilon - Q^2
 \end{eqnarray}
and we introduced $\eta^{\rm TE}=\eta c/(i\omega \ell)$ and $\eta^{\rm TM}=i\omega \epsilon \ell \eta$. The existence of a guided mode, which is understood as a mode propagating with a constant amplitude i.e. neither growing nor decaying, corresponds to the existence of an eigenvalue $Q$ such that $q$ is real {\em in spite of the presence of absorption and gain}.

Both problems (\ref{SL_TE}), and (\ref{SL_TM}) have eigenmodes of the form
\begin{equation}
\label{eigen}
\phi_{2n}^\alpha = \cos ({Q_{2n}^\alpha} Y)\quad\mbox{and} \quad  \phi_{2n+1}^\alpha = \sin (Q_{2n+1}^\alpha Y) 
\end{equation}
conveniently numbered by $n=0,1,...$, and referred below as cos- and sin-modes, respectively.
Corresponding sets of the eigenvalues are computed from the respective dispersion relations: 
\begin{subequations}
	\label{eigens}
\begin{equation}
	\label{eigens_TE}
 \cot Q_{2n} + \eta^\TE Q_{2n}=0,    \quad   \tan Q_{2n+1} - \eta^\TE Q_{2n+1}=0,
 \end{equation}
 \begin{equation}
 	\label{eigens_TM}
   Q_{2n} \tan Q_{2n} +\eta^\TM=0,    \quad  Q_{2n+1}\cot  Q_{2n+1} - \eta^\TM =0
\end{equation}
 \end{subequations}
for $\TE$-, and $\TM$-modes, respectively.  We notice that the above {mentioned} constraints on the sign of the real and imaginary parts of the surface impedance $\eta$, imply Re$(\eta^\TE)$, {Im$(\eta^\TE) <0$} and Re$(\eta^\TM)$, Im$(\eta^\TM) >0$.
  
\begin{figure}[ht!]
	\centering
	\includegraphics[width=1\columnwidth]{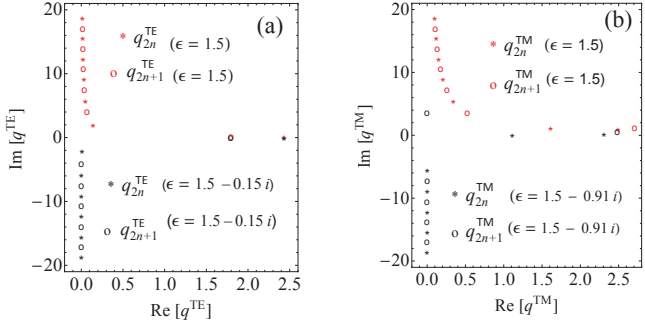}
	\caption{(Color online) Real {\em vs} imaginary parts of the dimensionless propagation constants, obtained for the wavenumbers  ${Q}$ in the range Re$\,{Q}\in[0,20]$, Im$\,{Q}\in[-20,20]$. (a) TE modes for ${\eta}^\TE=-1.65 - 2.06 i$ and $\epsilon_i = 0$ (red) and $\epsilon_i= - 0.15$ (black). (b)  TM modes for ${\eta}^\TM=1.65 +1.86 i$, and $\epsilon_i=0$ (red) and $\epsilon_i= - 0.91$ (black). In both panels $\epsilon_r = 1.5$, and $\lambda=3.1\ell$ have been set. Asterisks and empty circles represent cos-- and sin--modes, respectively.}
	\label{fig:two}
\end{figure}

In Fig.~\ref{fig:two}, we show typical examples of the ``distributions" of the propagation constants   on the complex plane $q=q_r + i q_i$ for the TE and TM modes. These are  inhomogeneous distributions with respect to  $q_r$. In both cases there exist a few modes with relatively large propagation constants and low decrements. These are "almost" guided modes. All other modes have large $q_i$ and small $q_r$. These modes are either absorbed by the boundaries at short distances, or amplified if the dielectric is active. All the modes are characterized by the energy flow from the waveguide centre to its boundaries.  In Fig.~\ref{fig:two} (a) one observes two modes, with low decrements, stem from the two guided modes of the passive waveguide with zero impedance.  

  To understand better the mode ``dynamics" upon change of the {boundary} surface properties, we consider a passive non-dissipative dielectric ($\epsilon_i=0$)  when the cladding is made of two almost perfect metals, i.e. when $|\eta^{TE}|\ll 1$ (the case of TM modes is treated similarly). For $n$ not too large, the eigenvalues $Q_{n}^{\TE}$ can be searched in the form  of the expansion $Q_{n}^{\TE}=Q_n+\eta^{TE}Q_n^{(1)}+\cdots $ where $Q_n=(n+1)\pi/2$ are the eigenvalues of the lossless waveguide with ideal metallic cladding (notice that because of the smallness of the impedance the system now is far from eventual EPs, which are discussed below). The other terms of this perturbation expansion are found by the straightforward algebra (not shown here) and lead to the approximation    {$Q_{n}^{\TE}=\pi(n+1)/[2(1- \eta^\TE)]+\mathcal{O}(n^3\eta^3)$} where $n=0,1,...$. 
  
  Let now in the limiting case, i.e. at  $\eta^\TE=0$, there exist $N$ guided modes i.e. $Q_N
 < \ell k_0\sqrt{\epsilon}=K<Q_{N+1}$. Then one computes 
 \begin{eqnarray}
 \begin{array}{ll}
 q_n\approx\sqrt{K^2-Q_n^2}-Q_n^2\eta^\TE/\sqrt{K^2-Q_n^2} &  \mbox{for $n\leq N$}
 \\
 q_n\approx i\sqrt{Q_n^2-K^2}+iQ_n^2\eta^\TE/\sqrt{Q_n^2-K^2} &  \mbox{for $n> N$}
 \end{array}
   \end{eqnarray}
 Taking into account that Re$(\eta^\TE)$, {Im$(\eta^\TE) <0$} we conclude that weak boundary absorption results in weak decay of modes guided by the conservative waveguide, but at the same time   {\em suppresses} the mode decay above the cut-off.  

\begin{figure}[ht!]
	\centering
	\includegraphics[width=1\columnwidth]{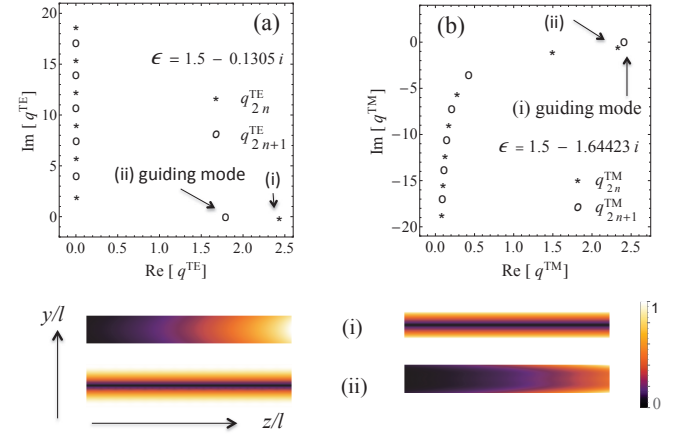}
	\caption{(Color online) Eigenvalue distribution showing guided mode. (a) TE modes for $\epsilon = 1.5 - 0.1305 i$ (b) TM modes for $\epsilon = 1.5 - 1.64423 i$. Lower panel:  absolute value of the TE (left) and TM (right) field evolutions for the modes indicated by (i), (ii) in (a) and (b), respectively. }
	\label{fig:guiding}
\end{figure}

The distributions of wavenumbers are strongly affected by the gain of an active dielectric: even relatively small gain may result in amplification of large (and even all, verified numerically) modes. Quite counter-intuitively, when gain increases namely decaying modes (rather than the two almost-propagating modes) become the amplifying ones, showing relatively weak effect on the modes with large $q_r$ and small decrements (cf. black and red sets of points in Fig.~\ref{fig:two}).  Moreover, larger increments are acquired by the modes having larger decrements. The described transition between the decaying and amplifying modes is narrow with respect to the gain $\epsilon_i$: all the modes which are decaying in Fig.~\ref{fig:two} (a) for $\epsilon_i< 0.0612$ become amplifying already for  $ \epsilon_i > 0.144$. Thus by applying homogeneous pump one can selectively amplify modes at the waveguide output. 
When ``shifting down" under the effect of the gain, each eigenvalue (in Fig.~\ref{fig:two}) crosses  the real axis, and at that instant the mode becomes guided. As an example consider  $\lambda=3.1\ell$ and $\epsilon = 1.5 - 0.1305 i$ for TE modes and $\epsilon = 1.5 - 1.64423 i$ for TM modes.  In both  cases there exists one guiding mode and all other modes are either decaying or amplifying [Fig.~\ref{fig:guiding}]. 

The non-monotonous dependence of the increment on the mode number naturally leads to the question about existence of the exceptions points (EPs), where two (or more) of the eigenstates coalesce with simultaneous coalescence of the respective eigenvectors. An EP can be found numerically using the approach as follows.  Denote by $f(Q,\eta)$ the left hand side of any of the dispersion relations in (\ref{eigens}), which now takes the form $f(Q,\eta)=0$. Suppose that it has a double root at $(\tilde{Q},\teta)$. That is at this point
$
f(\tQ,\teta) =0, 
$
$
\left. f_Q\right|_{(\tQ,\teta)} =0, 
$
and 
$
\left. f_{QQ}\right|_{(\tQ,\teta)} \ne  0, 
$
Now, using the Taylor expansion of the function $f(Q,\eta)$ in the vicinity of $(\tQ,\teta)$,
 one obtains
 \begin{equation}
f(Q,\eta) =  (\eta- \teta)f_\eta + (1/2) (Q- \tQ)^2 f_{QQ} +\cdots=0,
\end{equation} 
where all derivatives are evaluated at the double root. Thus in the leading order
$Q \approx \tQ \pm \sqrt{2  f_\eta/ f_{QQ}} \cdot \sqrt{ \teta - \eta},$
 implying that the eigenvalue $Q$ has a square root branch point at the double root $(\tQ,\teta)$ provided $\left. f_\eta\right|_{(\tQ,\teta)} \neq 0$.  Thus $(\tQ,\teta)$ defines the eigenvalue and the surface impedance corresponding to the EP. 
 
 We performed search of the EPs for both types of the modes according to the described algorithm. EPs for TE modes have not been found for the range of the parameters specified above (i.e. for specific signs of the real and imaginary parts of the  surface impedance). This does not rule out the existence of EPs for more sophisticated properties of cladding, say fabricated from meta-surfaces with negative permeability. Meantime we have found EPs for TM modes. Below we concentrate on this latter case assuming that the dielectric is passive (i.e. $\epsilon_i=0$). 
 
 \begin{figure}[ht!]
	\centering
	\includegraphics[width=1\columnwidth]{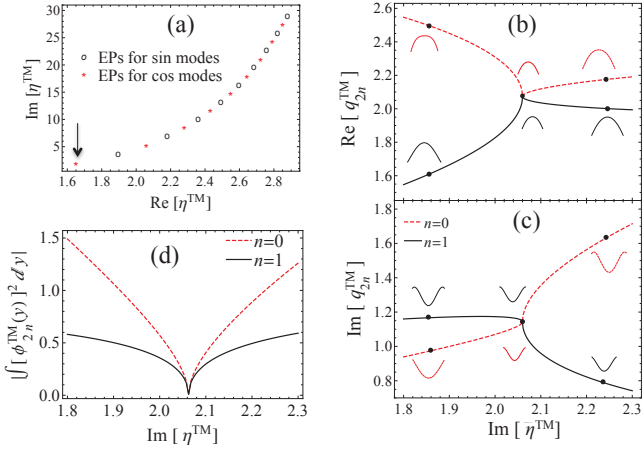}
	\caption{ (a) First few EPs are shown in the complex ${\eta}^{TM}$-plane. (b) and (c) respectively displays the real and imaginary parts of the eigenvalues {\it vs} Im$[\eta^{TM}]$ in the vicinity of first EP. (d) Numerically computed integral $I_{2n}$ is plotted in the vicinity of first EP. Here we set Re$[{\eta}^{TM}] = 1.6505$, $\epsilon = 1.5$, and $\lambda = 3.1 \ell$. In the inset of the panel (b) and (c) are the real and imaginary parts of the eigenfunctions, $\phi(y)$, at some selected eigenvalues (indicated by black dot). }
	\label{fig:four}
\end{figure}
 
The present setting possesses infinitely many such EPs {for TM modes,} some of which are shown in Fig.~\ref{fig:four}(a). The first of them, indicated by an arrow in the figure, corresponds to $\teta^{TM} = 1.65061 + 2.05998 i $ and $\tq^{TM} = 2.07346 + 1.14313 i$  {($\tQ = 2.1062 - 1.12536 i$)}, where $\tq$ is defined by (\ref{Q}) with $Q\to\tQ$, and used  $\epsilon = 1.5$ and $\lambda=3.1\ell$.  The real and imaginary parts of the eigenvalues in the vicinity of this point is shown in Fig.~\ref{fig:four} (b)  and (c), respectively. While both coalescing modes are decaying, however weaker absorption ``below" the EP, at Im($\eta^\TM)<$ Im($\teta$), is observed for $n=1$ mode, while ``above"  EP at Im($\eta^\TM) > $Im($\teta$) the mode $n=0$ decays more slowly. While in the points of coalescence the modes coincide, their profiles become distinguishable as $\eta^\TM$ varies (in panels (b) and (c) the mode profiles are shown for a few points of the impedance).  This means that at the output of sufficiently long waveguide, there will be observed slightly different fields patterns depending on the imaginary part of the surface impedance of the cladding, i.e. the effect of the {\em mode exchange} (which is a well known phenomenon and experimentally registered in microwave cavities~\cite{DGHHHRR}, and in waveguides~\cite{Doppler2016}).  

One of the salient feature of EPs is that the corresponding modes become self-orthogonal at these points, i.e. the integral $I_{2n} = | \int  \phi_{2n}^2 (Y) dY |$ becomes zero for the $n$-th coalescent eigenmode. To verify this property in Fig.~\ref{fig:four}(d) we show numerically evaluated integral in the vicinity of the first EP where the imaginary part of the surface impedance plays the role of parameter. At the EP the integral indeed vanishes. 

\begin{figure}[ht!]
	\centering
	\includegraphics[width=1\columnwidth]{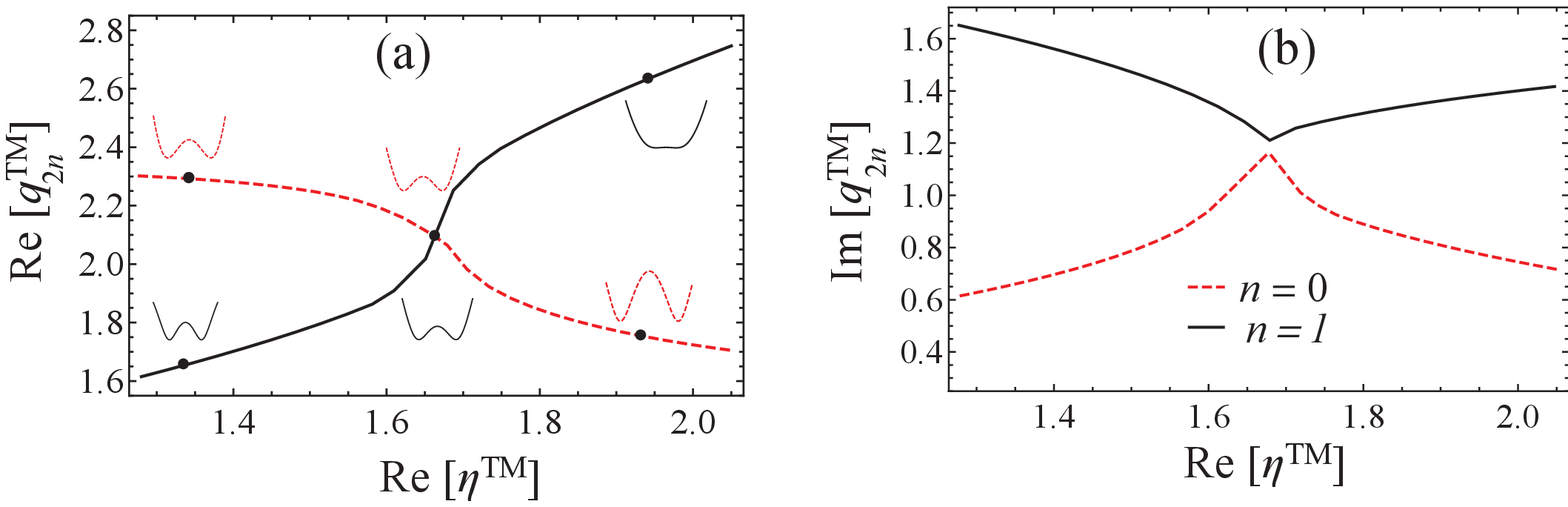}
	\includegraphics[width=.99\columnwidth]{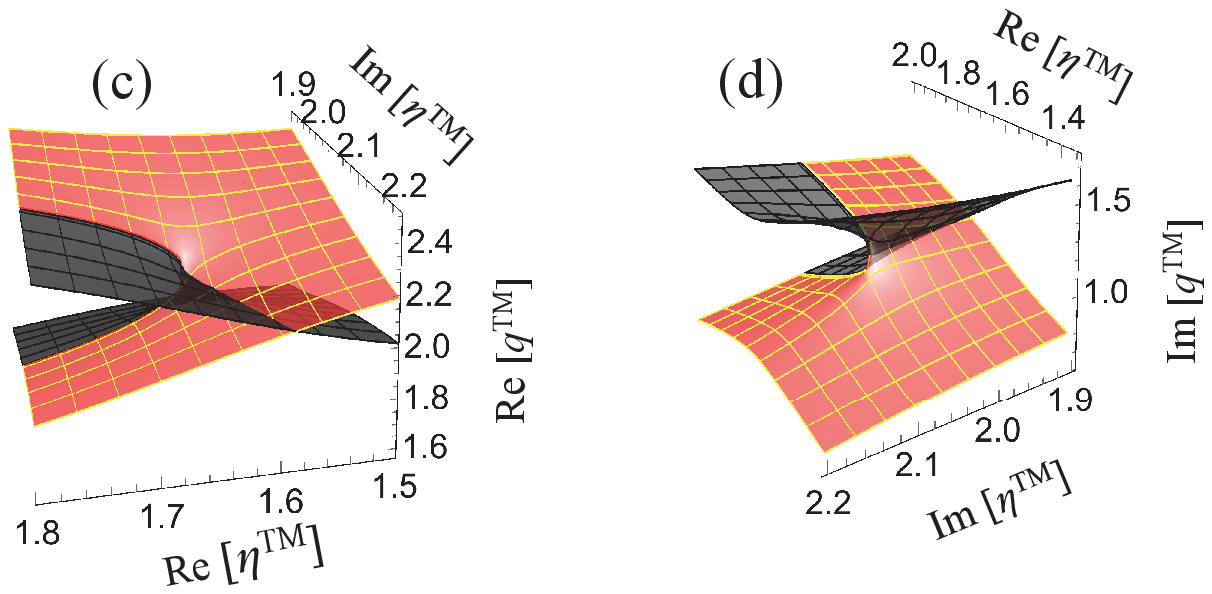}
	\caption{(Color online) (a) and (b) respectively display the real and imaginary parts of the eigenvalues  {\it vs}  Re$[\eta^{TM}]$ in the vicinity of the first EP. Here we set Im$[\eta^{TM}] = 2.0592$. In the inset of (a) are the absolute value of the eigenfunction $\phi(y)$ at the eigenvalues indicated by black dot.  The Riemann sheet structures near first EP are shown in the lower panels.} 
	\label{fig:five}
\end{figure}

 Furthermore, in Figs.~\ref{fig:four} and \ref{fig:five} we observe that two coalescent eigenmodes ``exchange" their properties when a control parameter crosses the EP. By increasing Im$[\eta^{TM}]$ one achieves that the first mode acquires significantly weaker absorption than the zero mode [Fig.~\ref{fig:four}]. By increasing Re$[\eta^{TM}]$ above the EP one makes the zero mode faster that the first mode [Figs.~\ref{fig:five}]. 	As a consequence, a certain input mode may lead to completely different output depending on the impedance.  The complete panorama of mode transformations, when both real and imaginary parts of the impedance are changed can be withdrawn from the Riemann surfaces shown in the lower panels of Fig.~\ref{fig:five}.
 
In conclusion, we have reported the possibility of selective mode guiding, switching between modes, and amplification, as well as the existence of EPs, in a planar waveguide with impedance boundary and passive or active dielectric filling. It is shown that by using active dielectric, pumped homogeneously and bounded by two metallic plates (or other absorbing metasurfaces with nonzero impedance), it is possible to amplify and guide modes selectively. The EPs found for TM modes, can be observed in experiments by tuning the transverse dimension of the waveguide or by changing the characteristics of cladding.
 Extending analysis beyond the approach based on the impedance boundary conditions, including, say, plasmonic modes which can be excited in cladding, may bring new possibilities for such control. The effect of topological properties of the EPs on the nontrivial light propagation~\cite{EP5,EP6}, also merits further thorough investigation.

\vspace{.1cm}
 {\it Acknowledgement.}  The research of B.M.  is supported by the People Programme (Marie Curie Actions) of the European Union's Seventh Framework Programme (FP7/2007-2013) under REA grant No. [291734].

\end{document}